\documentclass[prd,aps,twocolumn,amsmath,amssymb,floatfix,nofootinbib]{revtex4}

\usepackage{amsfonts,graphicx,amsfonts,multirow,epsfig,amsmath}
\usepackage[dvipdfm,colorlinks=true, citecolor=blue, linkcolor=blue, urlcolor=blue]{hyperref}
\usepackage{color,xcolor}

\newcommand{\ds}{\displaystyle}
\newcommand{\DS}[1]{/\!\!\!#1}

\allowdisplaybreaks

\begin{document}

\title{Revisiting the production of $J/\psi+\eta_c$ via the $e^+e^-$ annihilation within the QCD light-cone sum rules}

\author{Long Zeng$^1$}
\author{Hai-Bing Fu$^{1,4,}$}
\email{fuhb@cqu.edu.cn}
\author{Dan-Dan Hu$^1$}
\author{Ling-Li Chen$^1$}
\address{$^{\it 1}$Department of Physics, Guizhou Minzu University, Guiyang 550025, P.R. China}
\author{Wei Cheng$^{2,3}$}
\email{chengwei@itp.ac.cn}
\address{$^{\it 2}$Institute of Theoretical Physics, Chinese Academy of Sciences,  P.O.Box 2735, Beijing 100190,  P.R. China}
\address{$^{\it 3}$CAS Key Laboratory of Theoretical Physics, Institute of Theoretical Physics, Chinese Academy of Sciences, Beijing 100190, P.R. China}
\author{Xing-Gang Wu$^4$}
\email{wuxg@cqu.edu.cn}
\address{$^{\it 4}$Department of Physics, Chongqing University, Chongqing 401331, P.R. China}

\date{\today}

\begin{abstract}

We make a detailed study on the typical production channel of double charmoniums, $e^+e^-\to J/\psi+\eta_c$, at the center-of-mass collision energy $\sqrt{s}=10.58$ GeV. The key component of the process is the form factor $F_{\rm VP}(q^2)$, which has been calculated within the QCD light-cone sum rules (LCSR). To improve the accuracy of the derived LCSR, we keep the $J/\psi$ light-cone distribution amplitude up to twist-4 accuracy. Total cross sections for $e^+e^-\to J/\psi+\eta_c$ at three typical factorization scales are $\sigma|_{\mu_s} = 22.53^{+3.46}_{-3.49}~{\rm fb}$, $\sigma|_{\mu_k} = 21.98^{+3.35}_{-3.38}~{\rm fb}$ and $\sigma|_{\mu_0} = 21.74^{+3.29}_{-3.33}~{\rm fb}$, respectively. The factorization scale dependence is small, and those predictions are consistent with the BABAR and Belle measurements within errors.

\end{abstract}

\maketitle

\section{Introduction}

Double charmonium production at the $B$-factories has attracted large attention of experimentalists and theorists for a long time. At the beginning of this century, total cross section of $e^+e^-\to J/\psi+\eta_c$ at the  center-of-mass collision energy $\sqrt{s}=10.58$ GeV was firstly reported by the Belle Collaboration, $\sigma(e^+e^-\to J/\psi+\eta_c)\times{\cal B}_{\geq 4} = 33.0{^{+7.0}_{-6.0}}\pm 9.0~{\rm fb}$ with ${\cal B}_{\geq 4}$ being the branching ratio of $\eta_c$ into four or more charged tracks~\cite{Abe:2002rb}, which was update to $\sigma(e^+e^-\to J/\psi+\eta_c)\times{\cal B}_{\geq 2} = 25.6\pm2.8\pm 3.4~{\rm fb}$~\cite{Abe:2004ww}. Lately, the BABAR Collaboration issued their measured value $\sigma(e^+e^-\to J/\psi+\eta_c)\times{\cal B}_{\geq 2} = 17.6\pm2.8^{+1.5}_{-2.1}~{\rm fb}$~\cite{Aubert:2005tj}. Those measurements have severe discrepancy with the leading-order (LO) predictions based on the nonrelativistic QCD (NRQCD) factorization theory, which are within the range of $2.3\sim 5.5~{\rm fb}$~\cite{Braaten:2002fi, Liu:2002wq, Hagiwara:2003cw}. By including large and positive next-to-leading-order (NLO) contributions~\cite{Zhang:2005cha}, a larger total cross section $\sigma= 18.9~{\rm fb}$ by choosing the renormalization scale around $2-3$ GeV has been obtained, which is improved as $\sigma= 17.6 ^{+8.1}_{-6.7}~{\rm fb}$~\cite{Bodwin:2007ga} by further including relativistic corrections. A recent scale-invariant NRQCD prediction has been given in Ref.\cite{Sun:2018rgx} by applying the principle of maximum conformality (PMC)~\cite{Brodsky:2011ta, Brodsky:2012rj, Mojaza:2012mf, Brodsky:2013vpa}, which gives $\sigma= 20.35^{+3.5}_{-3.8}~{\rm fb}$, where the uncertainties are squared averages of the errors due to uncertainties from the charm-quark mass and the quarkonium wavefunction at the origin. Thus, it could be treated as another successful application of NRQCD.

The total cross-section of $e^+e^-\to J/\psi+\eta_c$ has also been studied by using the light-cone formalism~\cite{Ma:2004qf, Bondar:2004sv, Braguta:2005kr, Bodwin:2006dm}. Within the light-cone formalism, the amplitude of the process can be factorized as the perturbatively calculable short-distance part and the non-perturbative light-cone distribution amplitudes (LCDAs), which results in $\sigma= 14.4^{+11.2}_{-9.8}~{\rm fb}$~\cite{Braguta:2008tg}. The electromagnetic form factor $F_{\rm VP}(q^2)$ dominates the light-cone formalism, which can be calculated by using the QCD light-cone sum rules (LCSR). In Ref.\cite{Sun:2009zk}, after applying the operator production expansion (OPE) near the light cone and taking the $\eta_c$ leading-twist LCDA into account, the authors obtained a large factorization scale dependent total cross-section. By choosing the factorization scale as $\mu_s = 5.00{\rm GeV}$, the total cross-section is $\sigma|_{\mu_s} = 25.96\pm0.55~{\rm fb}$; and by setting the factorization as $\mu_k = 3.46{\rm GeV}$, the total cross-section changes to $\sigma|_{\mu_k} = 13.08\pm0.32~{\rm fb}$. A physical observable should be independent to the choice of factorization scale, and in the present paper, we shall adopt the LCSR approach to reanalyze the process and its factorization scale dependence.

The LCSR prediction should be independent to any choice of the correlator, an example for the QCD sum rules prediction of the $B$-meson constant $f_B$ under various choices of the correlator has been given in Ref.\cite{Wu:2010qt}. It is helpful to show whether other choices of correlator can also explain the data. As a new attempt, in the present paper, we shall adopt different correlator from Ref.\cite{Sun:2009zk} to do the LCSR calculation, in which the $J/\psi$ LCDAs other than the $\eta_c$ LCDAs shall be introduced. To improve the accuracy, we shall keep the $J/\psi$ LCDAs up to twist-4 accuracy, i.e., the resultant form factor $F_{\rm VP}(q^2)$ will contain $\phi_{2;J/\psi}^{\lambda}(x)$, $\phi_{3;J/\psi}^{\lambda}(x)$, $\phi_{4;J/\psi}^{\lambda}(x)$, $\psi_{4;J/\psi}^{\bot}(x)$ with $\lambda = (\|, \bot)$, which correspond to longitudinal and transverse distributions, respectively.

The remaining parts of the paper are organized as follows. In Sec.~\ref{section:2}, we present the calculation technology for dealing with the form factor $F_{\rm VP}(q^2)$ up to twist-4 accuracy within the LCSR approach. Our choices of the $J/\psi$ LCDAs shall also be given here. In Sec.~\ref{section:3}, the phenomenological results and discussions are presented. Section~\ref{section:4} is reserved for a summary.

\section{Theoretical framework}\label{section:2}

\subsection{Cross section for $e^{+}+e^{-}\to J/\psi +\eta_c$}

In this subsection, we give a brief review on how to calculate the cross-section of the process $e^{+}(p_1) +e^{-}(p_2) \to J/\psi(p_3) + \eta_c(p_4)$, which can be written as~\cite{Tanabashi:2018oca}
\begin{eqnarray}
\sigma &=& \frac{1}{4E_1 E_2 v_{\rm rel}}\int\frac{d^3 \vec{p}_3 d^3
\vec{p}_4} {(2\pi)^3 2E_3(2\pi)^3 2E_4}(2\pi)^4 \nonumber \\
&&\qquad\qquad\quad \times \delta^4 (p_1+p_2-p_3-p_4)|\overline{\cal M}|^2,
\end{eqnarray}
where $p_i=(E_i, \vec{p}_i)$ stands for the four-momentum of the initial or final particle, and the relative velocity between positron and electron, $v_{\rm rel}=|\vec{p}_1/E_1-\vec{p}_2/{E_2}|$. $|\overline{\cal M}|^2$ is the squared absolute value of the matrix element, where the color states and spin projections of the initial and final particles have been summed up and those of the initial particles have been averaged. The matrix element ${\cal M}$ can be written as
\begin{align}
&{\cal M} = i\int d^4 x \cr
&\quad \times \langle V P| T\big\{Q_c J^c_\mu (x) A^\mu (x), \bar e(0)\gamma_{\nu}e(0)A^{\nu}(0)\big\} |e^+e^-\rangle .\cr
\end{align}
Hereafter, to simplify the notation, we set $V=J/\psi$ and $P=\eta_c$. The $c$-quark electromagnetic current $J^{c}_{\mu}(x)=\bar{c}(x)\gamma_\mu c(x)$. Then, we obtain
\begin{align}
|\overline{\cal M}|^2 = 2Q_c^2 |F_{\rm VP}(q^2)|^2
\frac{\sqrt{2|{\bf p}|}}{4s}\left[1+\cos^2\theta\right],
\end{align}
where $\theta$ is the scattering angle, $Q_c = 2/3$ is the charge of $c$-quark, $s=-q^2=(p_1+p_2)^2$ or $(p_3+p_4)^2$, $|\bf p|$ is the magnitude of the three-momentum of one of the final-state mesons in the center-of-mass frame.

The form factor $F_{\rm VP} (q^2)$ is defined through the following matrix element~\cite{Bondar:2004sv}
\begin{eqnarray}
\langle J/\psi(p_3,\lambda), \eta_c(p_4)|J^V_\mu|0\rangle = \varepsilon_{\mu \nu\alpha\beta}\tilde\epsilon^{*(\lambda)\nu} p_3^\alpha p_4^\beta F_{\rm VP}(q^2),
\label{Eq:fvpdef}
\end{eqnarray}
where $\epsilon^\nu$ is the polarization vector of $J/\psi$. Neglecting the spin-flitting effects, we have $m_{\eta_c}=m_{J/\psi}$, and the cross section becomes
\begin{eqnarray}\label{Eq:crosssection}
\sigma &=&\frac{\pi \alpha^2 Q_c^2}{6} \left(1-\frac{4m_{J/\psi}^2}{s}\right)^{3/2} |F_{\rm VP}(q^2)|^2.
\end{eqnarray}

\subsection{The form factor $F_{\rm VP} (q^2)$ within the QCD LCSR}

To derive the form factor $F_{\rm VP} (q^2)$ within the QCD LCSR approach, we start with the following two-point correlation function (correlator)
\begin{eqnarray}
\Pi _{\mu\nu}(p,q) &=& i\int d^4 x e^{iq \cdot x}\langle V (p,\lambda )|T\{ J_\mu ^V(x),J_\nu^A(0)\} |0\rangle, \nonumber \\
\label{Eq:correlator2}
\end{eqnarray}
where $q$ and $p$ are four-momentum of the virtual photon and $J/\psi$. The current $J_\nu^A(x)= \bar c(x)\gamma_\nu \gamma_5 c(x)$ is the $c$-quark axial-vector current.

On the one hand, we deal with the hadronic representation of the correlator. It can be calculated by inserting a complete set of the intermediate hadronic states into the correlator, e.g.
\begin{align}
\Pi_{\mu\nu}(p,q) &= \frac{\langle V (p,\lambda)|J_\mu^V(0)|P(p-q)\rangle \langle P(p-q)|J_\nu^P(0)|0\rangle}{m_P^2 - (p-q)^2}
\nonumber\\
&+ \frac1{\pi}\int_{s_0}^\infty ds \frac{{\rm Im} \Pi_{\mu \nu}}{s - (p - q)^2}~,
\label{Eq:HadronicExpression2}
\end{align}
where $\epsilon^\nu$ is the polarization vector of $J/\psi$ and $s_0$ is the continuum threshold parameter, whose value could be set near the squared mass of the lowest vector charmonium state. The dispersion integration in Eq.\eqref{Eq:HadronicExpression2} contains the contributions from the higher resonances and the continuum states. The matrix element $\langle V (p,\lambda)|J_\mu^V(0)|P (p - q)\rangle$ and $\langle P (p-q)|J_\nu^A(0)|0\rangle$ are defined as
\begin{align}
&\langle V(p,\lambda)|J_\mu^V(0)|P(p-q)\rangle = \varepsilon_{\mu\nu\alpha\beta} \tilde\epsilon^{*(\lambda)\nu}q^\alpha p^\beta F_{\rm VP}(q^2), \label{Eq:FormFactor1}
\\[2ex]
&\langle 0|J_\nu^A(0)|P(p - q)\rangle  = if_P (p - q)_\nu, \label{Eq:DecayConstant1}
\end{align}
where $f_P$ is the $\eta_c$ decay constant. Inserting Eqs.(\ref{Eq:FormFactor1}, \ref{Eq:DecayConstant1}) into Eq.\eqref{Eq:HadronicExpression2}, we obtain
\begin{align}
\Pi_{\mu \nu }^{\rm Had}(p,q)&= \varepsilon_{\mu \nu \alpha \beta} \tilde\epsilon^{*(\lambda)\alpha} p^\beta \frac{ m_P^2 f_P F_{\rm VP}(q^2)}{{m_P^2 - {{(p- q)}^2}}} \nonumber\\[1ex]
& + \frac1 \pi \int_{s_0}^\infty  ds \frac{ F_{\mu\nu}(q^2) }{s-(p-q)^2}.
\end{align}

On the other hand, the correlator in the large space-like region, i.e. $(p+q)^2-m_c^2\ll 0$ with $q^2 \sim {\cal O}(1\;{\rm GeV})\ll m_c^2$ for the momentum transfer, corresponds to the $T$-product of quark currents near small light-cone distance $x^2\to 0$, which can be treated by operator product expansion (OPE) with the coefficients being pQCD calculable. For such purpose, we contract the two $c$-quark fields and write down a free $c$-quark propagator with gluon field $S^c(x,0) = \langle0|c_\alpha^i(x)\bar c_\beta^j(0)|0\rangle$ as follows~\cite{Fu:2020vqd, Fu:2020uzy}
\begin{align}
\langle0&|c_\alpha^i(x)\bar c_\beta^j(0)|0\rangle  = -i\int \frac{d^4k}{(2\pi)^4}e^{-ik\cdot x}\bigg\{\delta^{ij}\frac{\DS k + m_c}{m_c^2-k^2}
\nonumber\\[0.5ex]
&+g_s\int_0^1 dv~G^{\mu\nu}(vx)\,\left(\frac{\lambda}{2}\right)^{ij}\bigg[\frac{\DS k+m_c}{2(m_c^2 - k^2)^2}\sigma_{\mu\nu}
\nonumber\\[0.5ex]
&+ \frac1{m_c^2-k^2}vx_\mu\gamma_\nu\bigg]\bigg\}_{\alpha\beta}. \label{Eq:propagator}
\end{align}
Substituting Eq.\eqref{Eq:propagator} into the correlator, one needs to deal with the matrix elements of the nonlocal operators between vector meson and vacuum state, that is,
\begin{align}
&\langle V (p,\lambda)| \bar q_1(x) \sigma_{\mu\nu} q_2(0)|0\rangle = i f_V^\bot \int_0^1  du \tilde\epsilon^{iup\cdot x} \nonumber\\[1ex]
&\qquad \times \bigg\{ (\tilde\epsilon_\mu^{*(\lambda)}p_\nu  - \tilde\epsilon_\nu^{*(\lambda )}p_\mu) \bigg[ \phi_{2;V}^\bot (u) + \frac{m_V^2 x^2}{4} \phi_{4;V}^\bot(u) \bigg]
\nonumber\\
&\qquad + (p_\mu x_\nu  - p_\nu x_\mu) \frac{\tilde\epsilon^{*(\lambda )}\cdot x}{(p\cdot x)^2} m_{V}^2 \bigg[ \phi_{3;V}^\| (u) - \frac12 \phi_{2;V}^\bot(u)
\nonumber\\
&\qquad - \frac12 \psi_{4;V}^\bot(u) \bigg]
 + \frac12 ( \tilde\epsilon_\mu^{*(\lambda )}x_\nu  - \tilde\epsilon_\nu^{*(\lambda )}x_\mu )\frac{m_{V }^2}{p\cdot x}\bigg[\psi_{4;V}^ \bot(u)\nonumber\\
&\qquad - \phi_{2;V }^\bot (u) \bigg]\bigg\},
\label{Eq:DA1}
\end{align}
\begin{align}
&\langle V (p,\lambda )|\bar q_1(x)\gamma_\mu q_2(0)|0\rangle  = m_V f_V^\| \int_0^1 d u e^{iup \cdot x}
\nonumber\\
&\qquad\times\bigg\{ \tilde\epsilon_\mu ^{*(\lambda )}\phi_{3;V}^\bot (u)
+\frac{\tilde\epsilon^{*(\lambda )} \cdot x}{p\cdot x} p_\mu \bigg[ {\phi _{2;V }^\parallel (u) + \phi _{3;V }^ \bot (u)} \bigg]
\nonumber\\
&\qquad + \frac{{{\tilde\epsilon^{*(\lambda )}} \cdot x}}{{(p \cdot x)}}{p_\mu }\frac{{m_{V }^2{x^2}}}{4}\phi _{4;V }^\parallel (u)
- \frac12 x_\mu \frac{{{\tilde\epsilon^{*(\lambda )}} \cdot x}}{{{{(p \cdot x)}^2}}}m_V^2
\nonumber\\
&\qquad \times \bigg[ {\psi _{4;V }^\parallel (u) + \phi _{2;V }^\parallel (u) - 2\phi _{3;V }^ \bot (u)} \bigg] \bigg\},
\label{Eq:DA2}
\end{align}
and
\begin{align}
&\langle V (p,\lambda )|\bar q_1(x)i\gamma_\mu g G_{\alpha\beta}(vx)q_2(0)|0\rangle  = p_\mu (\tilde\epsilon_{\bot\alpha}^{*(\lambda )}p_\beta - \tilde\epsilon_{\bot\beta }^{*(\lambda)}p_\alpha)
\nonumber\\
&\qquad \times f_V^\| m_V \Phi_{3;V}^\| (v, p\cdot x) + (p_\alpha g_{\mu\beta}^\bot - p_\beta g_{\mu\alpha}^\bot)\frac{\tilde\epsilon^{*(\lambda )}\cdot x}{p\cdot x}
\nonumber\\
&\qquad \times f_V^\| m_V^3 \Phi_{4;V}^\| (v, p\cdot x) + {p_\mu }({p_\alpha }{x_\beta } - {p_\beta }{x_\alpha })\frac{\tilde\epsilon^{*(\lambda )}\cdot x}{p\cdot x}
\nonumber\\
&\qquad \times f_V^\| m_V^3 \Psi_{4;V}^\| (v,p\cdot x) + \ldots.\label{Eq:ThreeP}
\end{align}
The $J/\psi$ LCDAs $\phi_{2;V}^{\|,\bot}(u)$, $\phi_{3;V}^{\|,\bot}(u)$ and $\phi_{4;V}^{\|,\bot}(u)$/$\psi_{4;V}^{\bot}(u)$ stand for the two-particles twist-2, twist-3 and twist-4 ones, respectively; and the $J/\psi$ LCDAs $\Phi_{3;V}^\|(v,p\cdot x)$ and $\Phi_{4;V}^\|(v,p\cdot x)/\Psi_{4;V}^\|(v,p\cdot x)$ stand for the three-particles twist-3 and twist-4 ones, respectively.

Inserting the above LCDAs into the correlator \eqref{Eq:correlator2}, and completing the integration over $x$ and $k$, we can derive the OPE representation of the correlator. By equating both phenomenological and theoretical sides of the correlator and employ the usual Borel transform
\begin{eqnarray}
{\cal B}_{M^2} \Pi(q^2) = \mathop{\lim }\limits_{\scriptstyle - {q^2},n \to \infty \hfill\atop\scriptstyle - {q^2}/n = {M^2}\hfill} \frac{(-q^2)^{n+1}}{n!} \left(\frac d {dq^2} \right)^n \Pi (q^2),
\label{Eq:Borel}
\end{eqnarray}
the LCSR for the form factors $F_{\rm VP}(q^2)$ can be obtained, which reads
\begin{widetext}
\begin{eqnarray}
F_{\rm VP}(q^2) &=& \frac{m_{V}}{m_P^2 f_P}~\bigg\{ \int_0^1 du  e^{(m_P^2 - s(u))/M^2} \bigg\{ m_c m_{V} f_{V}^\bot ~ \bigg[ \frac1{um_{V}^2} \Theta(c(u,s_0)) \phi_{2;V}^\bot(u) - \frac{m_c^2}{u^3 M^4}
\nonumber\\
&\times & \tilde{\tilde\Theta}(c(u,s_0))\phi_{4;V}^\bot(u) - \frac 2{u^2 M^2} \tilde\Theta(c(u,s_0)) I_L(u) - \frac1{uM^2} \tilde\Theta (c(u,s_0))H_3(u) \bigg] + f_{V}^\|
\nonumber\\
&\times& \bigg[ \Theta(c(u,s_0))\phi_{3;V}^\bot (u) ~ +  \frac1 u\Theta(c(u,s_0))A(u)-m_{V}^2~\bigg( \frac{m_c^2}{u^3 M^4} \tilde{\tilde{\Theta}}(c(u,s_0)) + \frac1{u^2 M^2}
\nonumber\\
&\times& \tilde\Theta (c(u,s_0)) \bigg)~B(u) \bigg\} ~+~ f_V^\| \int {\cal D} \alpha_i \int dv~e^{(m_P^2-s(X))/M^2} ~\bigg[ m_V^2 (2v+ 1) \frac1{X M^2}
\nonumber\\
&\times &\tilde\Theta(c(X,s_0))~+~ (4v + 1)(m_V^2 - m_P^2 + q^2)~ \frac1{4X^2 M^2} ~\tilde\Theta(c(X,s_0)) \bigg] ~ \Phi_{3;V}^\| (\underline \alpha)\bigg\},
\label{Eq:FVP}
\end{eqnarray}
\end{widetext}
where $\alpha_i = (\alpha_1, \alpha_2, \alpha_3)$, $s(X) = [m_c^2 - \bar X(q^2 -Xm_{V}^2)]/X$ with $X=\alpha_1 + v \alpha_3$ and $\bar X = (1-X)$. The integration over $x$ can be done by transforming the $x_\mu$ in the nominator to $i\partial/\partial(up_\mu)$, or equivalently to $-i\partial/\partial q_\mu$, and make transformation
\begin{align}
\frac{1}{p\cdot x} \phi(u) \to -i\int_0^u dv\phi(v)\equiv-i\Phi(u).
\end{align}
The simplified distribution functions $I_L (u)$, $H_3(u)$, $A(u)$ and $B(u)$ are defined as:
\begin{align}
& I_L(u) = \int_0^u dv \int_0^v dw \bigg[\phi_{3;V}^\|(w) - \frac12\phi_{2;V}^\bot(w) - \frac12 \psi_{4;V}^\bot(w)\bigg],
\nonumber\\
& H_3(u) = \int_0^u dv \bigg[\psi_{4;V}^\bot (v) - \phi_{2;V}^\bot(v) \bigg],
\nonumber\\
& A(u) = \int_0^u dv \left[ \phi_{2;V}^\| (u) + \phi_{3;V}^\bot (u) \right],
\nonumber\\
& B(u) = \int_0^u dv \phi_{4;V }^\| (u).
\end{align}
The $\Theta(c(u,s_0))$ with $c(u,s_0) = us_0 -m_b^2 + \bar u q^2 - u\bar u m_{V}^2$ is the conventional step function, $\tilde\Theta [c(u,s_0)]$ and $\tilde{\tilde\Theta}[c(u,s_0)]$ take the following form
\begin{align}
&\int_0^1 \frac{du}{u^2 M^2} e^{-s(u)/M^2}\tilde\Theta(c(u,s_0))f(u)
\nonumber\\[1ex]
&\qquad= \int_{u_0}^1\frac{du}{u^2 M^2} e^{-s(u)/M^2}f(u) + \delta(c(u_0,s_0)),
\label{Theta1}\\
&\int_0^1 \frac{du}{2u^3 M^4} e^{-s(u)/M^2}\tilde{\tilde\Theta}(c(u,s_0))f(u)
\nonumber\\[1ex]
&\qquad= \int_{u_0}^1 \frac{du}{2u^3 M^4} e^{-s(u)/M^2}f(u)+\Delta(c(u_0,s_0)), \label{Theta2}
\end{align}
where
\begin{align}
\delta(c(u,s_0))&= e^{-s_0/M^2}\frac{f(u_0)}{{\cal C}_0},\nonumber\\
\Delta(c(u,s_0))&= e^{-s_0/M^2}\bigg[\frac{1}{2 u_0 M^2}\frac{f(u_0)} {{\cal C}_0}
\nonumber\\
&\left. -\frac{u_0^2}{2 {\cal C}_0} \frac{d}{du}\left( \frac{f(u)}{u{\cal C}} \right) \right|_{u = {u_0}}\bigg], \nonumber
\end{align}
${\mathcal C}_0 = m_b^2 + {u_0^2}m_V ^2 - {q^2}$ and $u_0$ is the solution of $c(u_0,s_0)=0$ with $0\leq u_0\leq 1$~\cite{Fu:2014pba}. Here we do not present the surface terms involving the three-particle LCDAs, since we have  found numerically that their contributions to the form factor are quite small and can be safely neglected.

\subsection{The $J/\psi$ LCDAs}

The important components for the form factor $F_{\rm VP} (q^2)$ are the gauge-independent and process-independent LCDAs, which can be derived from the wavefunction by integrating over the transverse components. For the $J/\psi$ LCDAs, we start from the following Brodsky-Huang-Lepage (BHL)~\cite{BHL} $J/\psi$ longitudinal/transverse twist-2 wavefunction,
\begin{align}
\psi_{2;J;\psi}^\lambda (x, {\bf k}_\bot) = \chi_{J/\psi} ({\bf k}_\bot) \psi_{2;J;\psi}^{\lambda,R} (x, {\bf k}_\bot),
\label{eq:WFSP}
\end{align}
where $\bf k_\bot$ stands for the transverse momentum, $\chi_{J/\psi} ({\bf k}_\bot)$ is the spin-space wavefunction which can be taken as the form  $\chi_{J/\psi} ({\bf k}_\bot) = \hat m_c / \sqrt{{\bf k}_\bot^2 + \hat m_c^2}$. The $\hat m_c = 1.8~{\rm GeV}$ is the constituent charm-quark mass~\cite{Sun:2009zk}. The spatial wavefunction $\psi_{2;J;\psi}^{\lambda,R} (x, {\bf k}_\bot)$ can be written as:
\begin{align}
\psi_{2;J;\psi}^{\lambda,R} (x, {\bf k}_\bot) = A_{J/\psi}^\lambda \exp\bigg[ - \frac1{8\beta_{J/\psi}^{\lambda~2}}~\frac{{\bf k}_\bot^2 + \hat m_c^2}{ x\bar x}\bigg],
\end{align}
where $\bar{x}=1-x$, $A_{J/\psi}^\lambda$ is normalization constant, and $\beta_{J/\psi}^\lambda$ is the harmonic parameter that dominantly determines the wavefunction transverse distributions. The LCDA can be obtained by integrating over the transverse momentum of the wavefunction, i.e.
\begin{align}
\phi_{2;J/\psi}^\lambda(x,\mu) = \frac{2\sqrt 6}{f_{J/\psi}^\lambda} \int_{|{\bf k}_\bot|^2 \leq \mu_0^2}~\frac{d^2{\bf k}_\bot}{16\pi^3}\psi_{2;J;\psi}^\lambda (x, {\bf k}_\bot). \label{eq:WF2DA}
\end{align}
where $\mu_0 = \hat m_c = 1.8~{\rm GeV}$~\cite{Sun:2009zk}. Then, we obtain
\begin{align}
\phi_{2;J/\psi}^\lambda(x,\mu)&=\frac{\sqrt{3}A_{J/\psi}^\lambda \hat m_c\beta_{J/\psi}^\lambda}{2\pi^{3/2}f_{J/\psi}^\lambda}\sqrt{x\bar{x}}\nonumber \\
&\times \bigg\{{\rm Erf}\bigg[\sqrt\frac{\hat m_c^2+\mu^2}{8\mu^2x\bar{x}}\bigg]-{\rm Erf}\bigg[\sqrt\frac{\hat m_c^2}{8\mu^8x\bar{x}}\bigg]\bigg\},
\label{eq:DABHL}
\end{align}
where $\lambda=\bot,\|$, and the error function ${\rm Erf}(x)=2\int_0^x e^{-t^2}dt /\sqrt{\pi}$. For the non-leading twist-3 wavefunction, we take the heavy quarkonium the light-front 1$S$-Coulomb form~\cite{Bondar:2004sv}
\begin{align}
\psi_{3; J/\psi}^{\rm Coulomb} \sim \bigg[\frac{{\bf k}_\bot^2+(1-4x\bar x)\hat m_c^2}{4x\bar x} + q_B^2\bigg]^{-2}
\end{align}
with $q_B$ is the Bohr momentum. After integrating with the transverse momentum $\bf k_\bot$, the fully expression can be written as
\begin{align}
\phi_{3; J/\psi}^{\lambda}(x,v^2) = c_i(v^2)\phi_{3;J/\psi}^{\lambda,{\rm Asy.}}(x) \bigg[\frac{x\bar x}{1-4x\bar x(1-v^2)}\bigg]^{1-v^2}
\end{align}
where the mean heavy quark velocity $v = q_B/\hat m_c\ll 1$, and we set $v^2 \simeq 0.30$~\cite{Wang:2020zbr} to do the numerical analysis. The twist-3 LCDAs are normalized to 1, i.e. $\int_0^1 \phi_{3; J/\psi}^{\lambda}(x,v^2) =1 $. Finally, the twist-3 LCDAs takes the following form:
\begin{align}
&\phi_{3; J/\psi}^{\|}(x) = 10.94\xi^2 \bigg[\frac{x\bar x}{1-2.8x\bar x}\bigg]^{0.70}, \cr
&\phi_{3; J/\psi}^{\bot}(x) = 1.67(1 + \xi^2) \bigg[\frac{x\bar x}{1-2.8x\bar x}\bigg]^{0.70},
\end{align}
where $\xi=2x-1$. The twist-3 LCDAs $\phi_{3; J/\psi}^{\lambda}(x)$ can also be derived from the twist-2 LCDAs $\phi_{2; J/\psi}^{\lambda}(x)$ by using the Wandzura-Wilczek approximation~\cite{Wandzura:1977qf, Ball:1997rj}. However we observe that the contribution of LCDAs from the end-point region $x\sim 0,1$ can not be effectively suppressed, leading to a unwanted large cross section. Thus we adopt the above light-front 1$S$-Coulomb form for the twist-3 wavefunction which is usually taken in the literature to deal with the double charmonium production.

Because the terms involving the twist-4 LCDAs are quite small in comparison to the twist-2 and twist-3 terms, so the uncertainties from the twist-4 LCDAs themselves could be negligible; thus we shall employ the twist-4 LCDAs $\phi_{4;J/\psi}^{\lambda}(x)$ and $\psi_{4;J/\psi}^{\bot}(x)$ without charm-quark mass effect that have been suggested by P. Ball and V.M. Braun~\cite{Ball:1998kk} to do the numerical calculation.

\section{Numerical Analysis}\label{section:3}

\subsection{Input parameters and the $J/\psi$ LCDAs}

To do the numerical calculation, we neglect the spin-flipping effect for the charmoniums and set the mass of $\eta_c$ or $J/\psi$ to be the same, $m_{\eta_c}=m_{J/\psi}=3.097~{\rm GeV}$~\cite{Tanabashi:2018oca}. As for the $J/\psi$ decay constant $f^\|_{J/\psi}$, we extract it from its leptonic decay width $\Gamma({J/\psi\to e^+e^-})$ by using the following relation~\cite{Hwang:1997ie}
\begin{eqnarray}
f_{J/\psi}^{\| 2} = \frac{3 m_{J/\psi} }{4\pi\alpha^2 c_{J/\psi}} \Gamma(J/\psi \to e^+e^-),
\end{eqnarray}
where $\alpha = 1/137$ and $c_{J/\psi} = 4/9$. Taking the PDG averaged value, $\Gamma(J/\psi \to e^+e^-) = 5.547(140)~{\rm KeV}$~\cite{Tanabashi:2018oca}, we obtain $f_{J/\psi}^\| = 416.2(53)~{\rm MeV}$. The transverse decay constant $f_{J/\psi}^\bot$ is taken as $0.410(10)~{\rm GeV}$~\cite{Becirevic:2013bsa}, and the $\eta_c$ decay constant $f_{\eta_c} = 0.453(4)$~\cite{Zhong:2014fma}.

The twist-2 wavefunction parameters $A_\lambda$ and $\beta_\lambda$ are fixed by two criteria:
\begin{itemize}
\item  The normalization condition of the twist-2 LCDA, i.e.
\begin{align}
\int \phi_{2;J/\psi}^{\lambda}(x,\mu)dx = 1.
\end{align}
\item  The Gegenbauer moment $a_n^{\lambda}$ and the twist-2 LCDA can be related via the following relation,
\begin{align}
~~~~~a_{n;J/\psi}^\lambda (\mu) = \dfrac{\ds\int_0^1 dx \phi_{2;J/\psi}^{\lambda} (x,\mu) C_n^{3/2}(2x-1)}{\ds\int_0^1 6x \bar x [C_n^{3/2}(2x-1)]^2}.  \label{eq:anphi}
\end{align}
One can derive the Gegenbauer moments $a_{n;J/\psi}^\lambda (\mu)$ of $\phi_{2;J/\psi}^{\lambda}$ by using their relationship to the moments, $\langle\xi_{n;J/\psi}^\lambda\rangle = \int_0^1 dx (2x-1)^n \phi_{2;J/\psi}^\lambda(x,\mu)$. More explicitly, we have
\begin{align}
\langle \xi_{2;J/\psi}^\lambda \rangle = \frac15\left( 1 + \frac{12}{7} a_{2;J/\psi}^\lambda\right).
\end{align}
The first moments of $\phi_{2;J/\psi}^{\lambda}$ has been calculated by Ref.\cite{Braguta:2007fh}, e.g., $\langle \xi_{2;J/\psi}^{\|} \rangle=0.070\pm0.0075$ and $\langle \xi_{2;J/\psi}^{\bot} \rangle =0.072\pm0.0075$ at the scale $\mu=1.2~{\rm GeV}$.
\end{itemize}

\begin{table}[htb]
\begin{center}
\caption{Two parameters of the $J/\psi$ longitudinal and transverse wavefunctions at the scale $\mu_0 = 1.8$ GeV.}\label{Tab:Abeta}
\begin{tabular}{c c c}\hline
~~~~~~$ $~~~~~~ &~~~~~~$A_{J/\psi}^\lambda$~~~~~~&~~~~~~$\beta_{J/\psi}^\lambda$~~~~~~        \\
\hline
$\phi_{2;J/\psi}^\|  $  & 458  & 0.682 \\
$\phi_{2;J/\psi}^\bot$  & 526  & 0.667 \\
\hline
\end{tabular}
\end{center}
\end{table}

The Gegenbauer moments at any other scale $a_{n;J/\psi}^\lambda (\mu)$ can be obtained via the QCD evolution. At the NLO accuracy, we have
\begin{align}
a_{n;J/\psi}^\lambda (\mu) &= a_{n;J/\psi}^\lambda (\mu_0) E_{n;J/\psi}^{\rm NLO}\cr
& + \frac{\alpha_s(\mu)}{4\pi} \sum_{k=0}^{n-2} a_{k;J/\psi}^\lambda (\mu_0) {\cal L}^{\gamma_k^{(0)} / (2\beta_0)} d_{nk}^{(1)}.
\end{align}
Here $\mu_0$ is the initial scale, $\mu$ is the required scale, and
\begin{align}
&E_{n;J/\psi}^{\rm NLO}  = {\cal L}^{\gamma_n^{(0)} / (2\beta_0)} \cr
&\qquad\times \bigg\{1+\frac{\gamma_n^{(1)}\beta_0 - \gamma_n^{(0)}\beta_1}{8\pi\beta_0^2}\big[\alpha_s (\mu) - \alpha_s(\mu_0)\big] \bigg\},
\end{align}
where ${\cal L} = \alpha_s(\mu)/\alpha_s(\mu_0)$, $\beta_0 = 11 - 2n_f/3$ and $\beta_1 = 102 - 38n_f/3$ with $n_f$ being the active flavor numbers. $\gamma_n^{(0)}$ stands for the anomalous dimensions to NLO accuracy, $\gamma_n^{(0)}$ is the diagonal two-loop anomalous dimension, and the mixing coefficients $d_{nk}^{(1)}$ with $k\leq n-2$ can be found in Ref.~\cite{Ball:2006nr}. For example, we present the central values for the input parameters of the $J/\psi$ longitudinal and transverse wavefunctions at the scale $\mu_0 = 1.8$ GeV in Table~\ref{Tab:Abeta}, where the LCDA moments are taken as $a_2^\|(\mu_0) = -0.321$ and $a_2^\bot(\mu_0) = -0.327$.

\begin{figure*}[htb]
\includegraphics[width=0.45\textwidth]{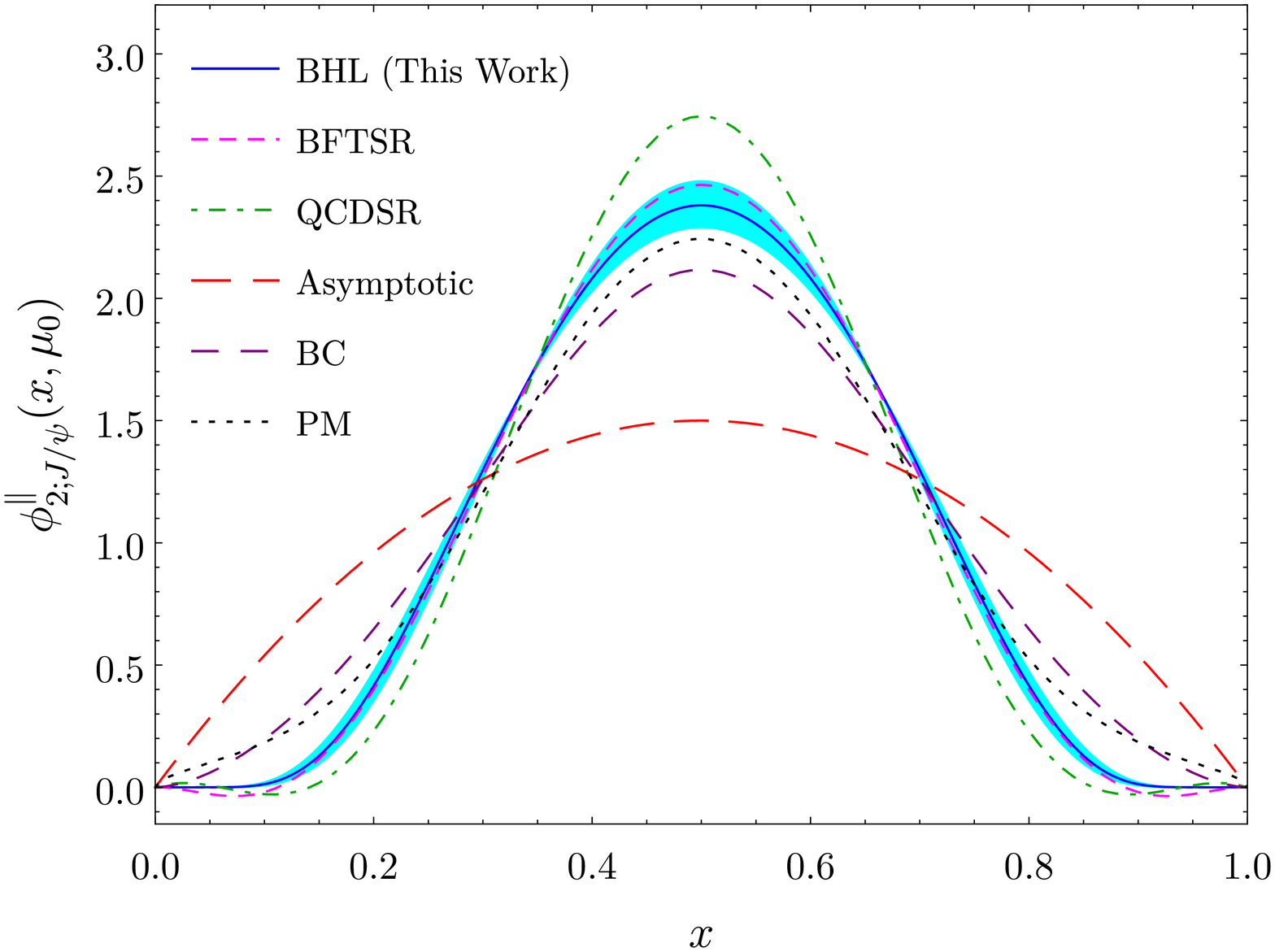}
\includegraphics[width=0.45\textwidth]{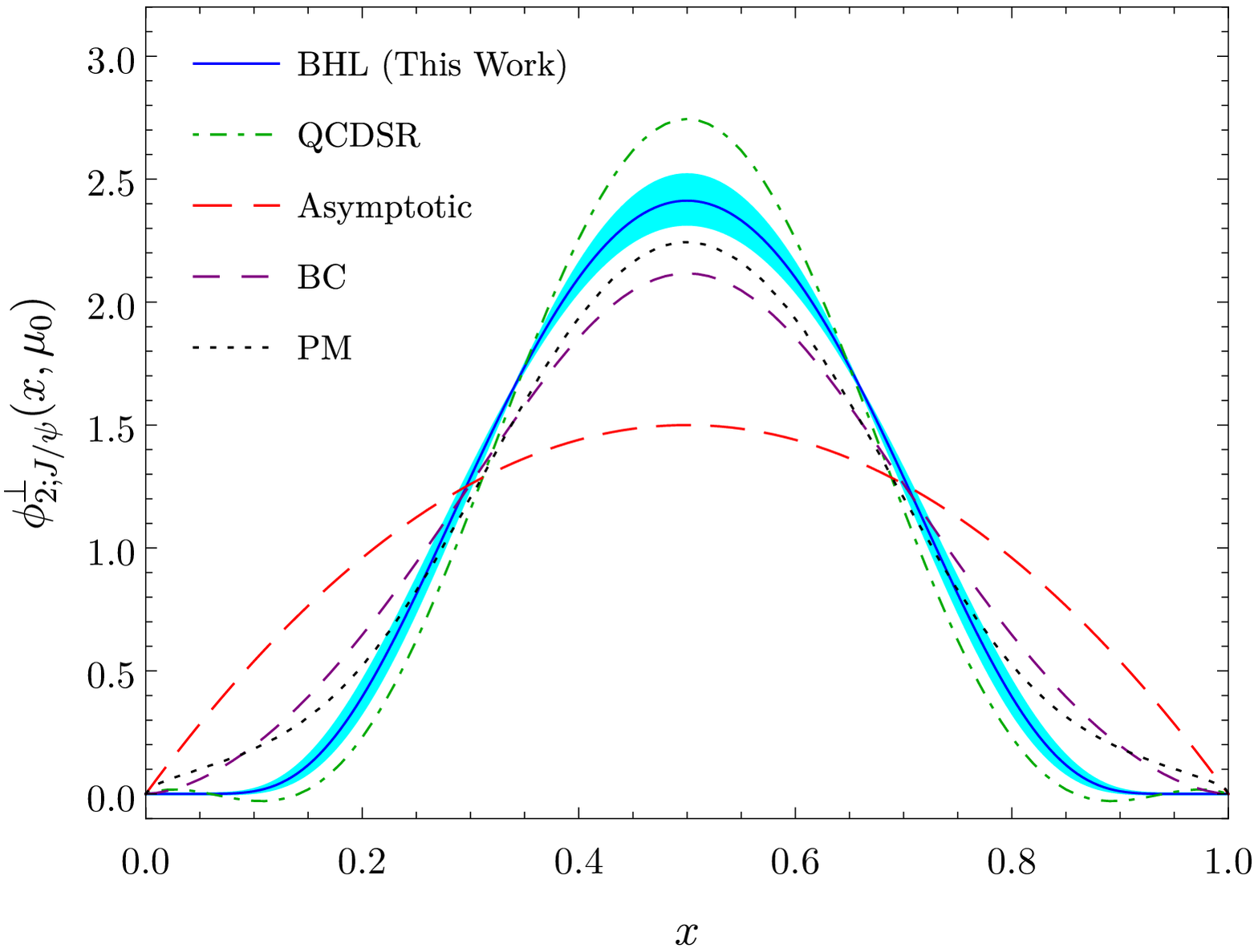}
\caption{The $J/\psi$ twist-2 LCDAs $\phi_{2;J/\psi}^\lambda(x,\mu)$ at the scale $\mu_0 = 1.8$ GeV, where $\lambda = (\|, \bot)$ stand for the longitudinal (Left diagram) and the transverse (Right diagram) parts, respectively. As a comparison, the asymptotic form, the BFTSR~\cite{Fu:2018vap}, the QCD SR~\cite{Braguta:2007fh}, the BC model~\cite{Bondar:2004sv}, and the potential model~\cite{Bodwin:2006dm} are also presented.}
\label{Fig:DALT}
\end{figure*}

Using those parameters, we present the $J/\psi$ longitudinal and transverse twist-2 LCDAs at the scale $\mu_0 = 1.8~{\rm GeV}$ in Fig.~\ref{Fig:DALT}. As a comparison, we also present the curves from various approaches in Fig.~\ref{Fig:DALT}, which are predicted by using the QCD sum rules~\cite{Braguta:2007fh}, the BFTSR~\cite{Fu:2018vap}, the model suggested by Bondar and Chernyak (BC)~\cite{Bondar:2004sv}, the model constructed from the potential model (PM)~\cite{Bodwin:2006dm} and the asymptotic form $\phi_{\rm asy.} = 6x\bar x$. Fig.~\ref{Fig:DALT} indicates that all the LCDA models prefer a single-peaked behavior, the BC and PM LCDAs are close in shape. Our present LCDA has a slightly sharper peak around $x\sim 0.5$ in agreement with the QCDSR and BFTSR, which has a stronger suppression around the ending point $x\sim 0, 1$. We find that the shape of $\phi_{2;J/\psi}^\|(x,\mu_0)$ LCDAs within uncertainties is almost the same as that of the BFTSR in the whole regions.

\subsection{$e^+e^-\to J/\psi +\eta_c$ cross section}

To derive the numerical results of $F_{\rm VP}(q^2)$, we need to fix the magnitudes of the effective threshold parameter $s_0$ and the Borel parameter $M^2$. As for $s_0$, we set $s_0 = 3.69^2 ~{\rm GeV^2}$~\cite{Eidemuller:2000rc} which is close to the squared mass of $\psi(2S)$. As for the Borel parameter $M^2$, we set it in the range $M^2\in[39, 41]~{\rm GeV}^2$. In this Borel window, not only the contributions of the higher resonance states and continuum states are greatly suppressed, but also the $M^2$-dependence is effectively suppressed~\cite{Sun:2009zk}.

As for the factorization scale $\mu$ of $e^{+} + e^{-} \to J/\psi + \eta_c$, to discuss the factorization scale dependence, in addition to the previously choice of $\mu=\mu_0$, we also take another two frequently choices to do our calculation, i.e. $\mu=\mu_k \approx \sqrt{k^2} \approx 3.46~{\rm GeV}$, which is determined by fixing the coupling constant $\langle\alpha_s(k^2)\rangle \approx 0.263$ and the mean value of $\langle Z_m^k\rangle\approx 0.80$~\cite{Bondar:2004sv}; and $\mu=\mu_s \approx \sqrt s/2 \approx 5~{\rm GeV}$~\cite{Braguta:2006wr}.

\begin{figure}[htb]
\includegraphics[width=0.48\textwidth]{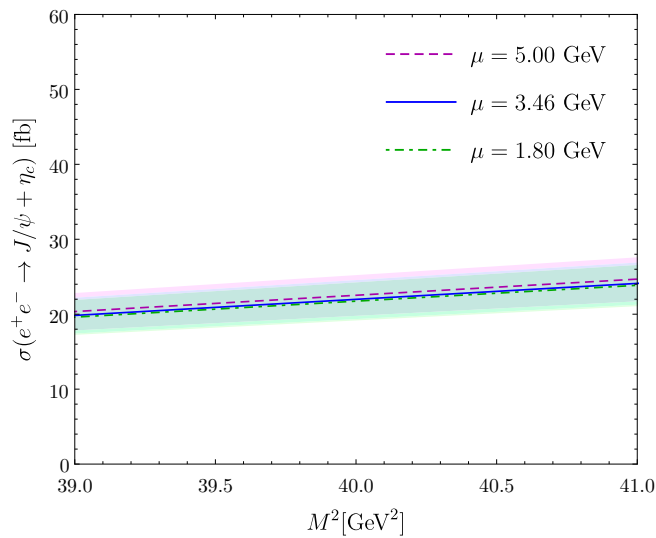}
\caption{Total cross-section of $e^{+}+e^{-}\rightarrow J/\psi+\eta_c$ at different factorization scale within the LCSR approach. The solid, dashed and dotted lines are the central values, which correspond to the $J/\psi$ distribution amplitude at the scale $\mu=\mu_0$, $\mu_k$ and $\mu_s$, respectively. The shaded bands are their errors from all inputs parameters.} \label{Fig:CSM2}
\end{figure}

Using those inputs together with the total cross section \eqref{Eq:crosssection}, we calculate the total cross-sections of $e^+e^-\to J/\psi +\eta_c$ under three different factorization scales, and we put their values versus the Borel parameter $M^2$ in Fig.~\ref{Fig:CSM2}. Fig.~\ref{Fig:CSM2} confirms that the total cross-section changes slightly within the allowable Borel widow, because the higher-twist terms are $1/M^2$-power suppressed.

\begin{table}[htb]
\caption{Uncertainties of the total cross section of $e^{+}+e^{-}\rightarrow J/\psi+\eta_c$ caused by the mentioned input parameters within the QCD LCSR approach.}
\label{Tab:sigmaall}
\begin{tabular}{lccc}
\hline
                                             & ~~~~~~~$\mu_s$~~~~~~~                   & ~~~~~~~$\mu_k$~~~~~~~                     &   $\mu_0$      \\ \hline
$\Delta M^2=\pm $                                 & $^{+2.16}_{-2.17}$       & $^{+2.13}_{-2.15}$          &   $^{+2.12}_{-2.13}$
\\[0.7ex]
$\Delta s_0=\pm $                                 & $^{+1.88}_{-1.79}$       & $^{+1.81}_{-1.72}$          &   $^{+1.77}_{-1.68}$
\\[0.7ex]
$\Delta m_c=\pm$                                 & $^{+1.80}_{+1.80}$       & $^{+1.73}_{-1.91}$          &    $^{+1.69}_{-1.87}$
\\[0.7ex]
$\Delta f_{\eta_c}=\pm$                          & $^{+0.30}_{-0.29}$       & $^{+0.29}_{-0.29}$          &     $^{+0.29}_{-0.29}$
\\[0.7ex]
$\Delta f_{J/\psi}^\|=\pm$                       & $^{+0.11}_{-0.11}$       & $^{+0.11}_{-0.11}$          &     $^{+0.11}_{-0.11}$
\\[0.7ex]
$\Delta f_{J/\psi}^\bot=\pm$                     & $^{+0.47}_{-0.47}$       & $^{+0.47}_{-0.46}$          &      $^{+0.47}_{-0.46}$
\\[0.7ex]
$\Delta \langle \xi_{2;J/\psi}^\| \rangle =\pm $   & $^{+0.01}_{-0.01}$       & $^{+0.01}_{-0.00}$          &      $^{+0.00}_{-0.00}$
\\[0.7ex]
$\Delta \langle \xi_{2;J/\psi}^\bot \rangle=\pm$ & $^{+0.36}_{-0.27}$       & $^{+0.20}_{-0.13}$          &      $^{+0.06}_{-0.03}$
\\[0.7ex]
\hline
\end{tabular}
\end{table}

To have a clear look at the errors coming from all the input parameters, we list the errors caused by each parameter in Table~\ref{Tab:sigmaall}. When discussing the error from one input parameter, all the other input parameters are set to be their central values. By adding up all the errors in mean square, our final LCSR predictions for the total cross-section of $e^{+}+e^{-}\to J/\psi+\eta_c$ at three typical factorization scales are
\begin{align}
&\sigma|_{\mu_s} = 22.53^{+3.46}_{-3.49}~{\rm fb}, \\
&\sigma|_{\mu_k} = 21.98^{+3.35}_{-3.38}~{\rm fb}, \\
&\sigma|_{\mu_0} = 21.74^{+3.29}_{-3.33}~{\rm fb}.
\end{align}
Those cross-sections are close to each other, indicating the factorization scale dependence is small. Thus by properly dealing with the QCD evolution effect, the LCSR predictions shall be slightly affected by different choice of factorization scale.

\section{Summary}\label{section:4}

\begin{figure}[htb]
\includegraphics[width=0.45\textwidth]{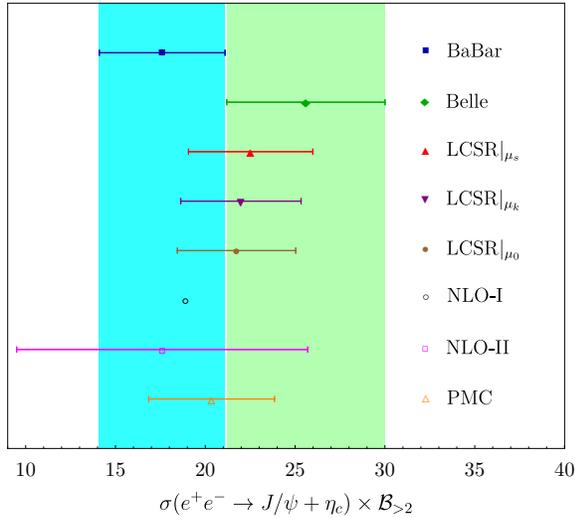}
\caption{Total cross section of $e^{+}+e^{-}\rightarrow J/\psi+\eta_c$ at different factorization scales within the LCSR approach. The marks represent the corresponding central values, and lines are the errors from the variation of all inputs parameters. As a comparison, the Belle data~\cite{Abe:2004ww}, the BaBar data~\cite{Aubert:2005tj}, the NLO NRQCD prediction (NLO-I)~\cite{Zhang:2005cha}, the NRQCD prediction with NLO radiative and relativistic corrections (NLO-II)~\cite{Bodwin:2007ga}, and the PMC NLO NRQCD prediction~\cite{Sun:2018rgx} are also presented.} \label{Fig:CSM2tot}
\end{figure}

In this paper, we have investigated the total cross-section for $e^+ e^- \to J/\psi+ \eta_c$ within the QCD LCSR approach. We put a comparison of total cross-section with other theoretical and experimental predictions in  Fig.~\ref{Fig:CSM2tot}. Fig.~\ref{Fig:CSM2tot} shows that our results are in consistent with the BaBar and Belle measurements and also the PMC NRQCD prediction within errors. Thus the LCSR approach also provides a helpful and reliable approach to deal with the high-energy processes involving charmoniums.

{\bf Acknowledgments}: We are grateful to Dr. Tao Zhong and Xu-Chang Zheng for helpful discussions and valuable suggestions. Hai-Bing Fu would like to thank the Institute of Theoretical Physics in Chongqing University for kind hospitality. This work was supported in part by the National Natural Science Foundation of China under Grant No.11765007, No.11947406 and No.11625520, the Project of Guizhou Provincial Department of Science and Technology under Grant No.KY[2019]1171, the Project of Guizhou Provincial Department of Education under Grant No.KY[2021]030 and No.KY[2021]003, the China Postdoctoral Science Foundation under Grant No.2019TQ0329 and No.2020M670476, and the Fundamental Research Funds for the Central Universities under Grant No.2020CQJQY-Z003.

\end{document}